\documentclass[twoside]{article}
\usepackage[margin=1in]{geometry}

\usepackage{microtype}
\usepackage{graphicx}
\usepackage{subfigure}
\usepackage{booktabs} 
\usepackage{natbib}
\usepackage{url}
\usepackage{algorithm}
\usepackage{algorithmic}
\usepackage{amsmath,amsthm,amsfonts,amssymb}
\usepackage{bbm}
\usepackage{comment}
\usepackage{color}
\usepackage{mathrsfs}
\usepackage{enumitem}
\usepackage{bm}
\usepackage{multirow}
\usepackage{booktabs}
\usepackage{makecell}
\usepackage{graphicx}
\usepackage{subfigure}
\usepackage{caption}
\usepackage{thmtools}
\usepackage{times}
\usepackage{dsfont}
\usepackage[dvipsnames]{xcolor}
\usepackage{natbib}

\usepackage{hyperref}
\usepackage[capitalise, noabbrev]{cleveref}
\hypersetup{
    colorlinks=true,
    linkcolor=blue,
    filecolor=magenta,      
    urlcolor=cyan,
    pdftitle={Choosing a Proxy Metric from Past Experiments},
    pdfpagemode=FullScreen,
    }
\newcommand{\hatproxvec}[1]{\mathbf{\hat{\Delta}}_{#1}^P}
\newcommand{\proxvec}[1]{\mathbf{\Delta}_{#1}^P}

\newcommand{\corr}{\mathrm{corr}}
\newcommand{\Cov}{\mathrm{Cov}}
\newcommand{\Var}{\mathrm{Var}}
\newcommand{\rref}{\mathrm{ref}}
\newcommand{\bepsilon}{\bm{\epsilon}}
\newcommand{\tstat}{\mathrm{tstat}}

\newcommand{\vect}[1]{\ensuremath{\mathbf{#1}}}

\newcommand{\mat}[1]{\ensuremath{\mathbf{#1}}}

\newcommand{\norm}[1]{\|{#1} \|}

\newcommand{\mE}{\mathds{E}}
\renewcommand{\Pr}{\mathds{P}}
\newcommand{\mR}{\mathds{R}}

\newcommand{\I}{\mat{I}}

\newcommand{\w}{\vect{w}}
\newcommand{\x}{\vect{x}}

\newcommand{\cD}{\mathcal{D}}

\newcommand{\cN}{\mathcal{N}}

\title{Choosing a Proxy Metric from Past Experiments}
\author{
Nilesh Tripuraneni\\
Google DeepMind\\
\small{\texttt{nileshtrip@google.com}} \hspace{-1em}
\and
Lee Richardson\\
Google\\
\small{\texttt{leerich@google.com}}
\and
Alexander D’Amour\\
Google DeepMind\\
 \small{\texttt{alexdamour@google.com}}
\and
Jacopo Soriano\\
Google\\
 \small{\texttt{jacoposoriano@google.com}}
\and
Steve Yadlowsky\\
Google DeepMind\\
 \small{\texttt{yadlowsky@google.com}}
}
\date{}

\usepackage{natbib}
\usepackage{graphicx}

\begin{document}

\maketitle

\begin{abstract}
    In many randomized experiments, the treatment effect of the long-term metric (i.e. the primary outcome of interest) is often difficult or infeasible to measure. Such long-term metrics are often slow to react to changes and sufficiently noisy they are challenging to faithfully estimate in short-horizon experiments. A common alternative is to measure several short-term proxy metrics in the hope they closely track the long-term metric -- so they can be used to effectively guide decision-making in the near-term. We introduce a new statistical framework to both define and construct an optimal proxy metric for use in a homogeneous population of randomized experiments. Our procedure first reduces the construction of an optimal proxy metric in a given experiment to a portfolio optimization problem which depends on the true latent treatment effects and noise level of experiment under consideration. We then denoise the observed treatment effects of the long-term metric and a set of proxies in a historical corpus of randomized experiments to extract estimates of the latent treatment effects for use in the optimization problem. One key insight derived from our approach is that the optimal proxy metric for a given experiment is not apriori fixed; rather it should depend on the sample size (or effective noise level) of the randomized experiment for which it is deployed.
    To instantiate and evaluate our framework, we employ our methodology in a large corpus of randomized experiments from an industrial recommendation system and construct proxy metrics that perform favorably relative to several baselines.
\end{abstract}

\section{Introduction}
Randomized controlled trials (RCTs) are the gold standard approach for measuring the causal effect of an intervention \citep{hernan2010causal}; however, designing and analyzing high-quality RCTs requires various considerations to ensure scientifically robust results. For example, an experimenter must clearly define the intervention, control, and choose a primary outcome for the study. In this work, we will assume that the intervention and control are clearly defined, and consider the problem of choosing a good primary outcome. A common approach is to choose the primary outcome to be a key metric which drives downstream decision-making. Such metrics are critical components in the decision-making pipelines of many large-scale technology companies \citep{Chen_Xin_2017, northstar2021} as well as used to guide policy decisions in economics and medicine \citep{NBERw26463, elliott2015surrogacy}. Unfortunately, direct measurement of such a metric can be impractical or infeasible. In many cases, they are long-term outcomes
observed with a significant temporal delay, making them slow to move (i.e. insensitive) in the short term, and inherently noisy. Moreover, they may be prohibitively expensive to query.

On the other hand, proxy metrics (or surrogates) that are easier to measure or faster to react are often available to use in lieu of the long-term outcome. For example in clinical settings, CD4 white-blood cell counts in blood serve as a surrogate for mortality due to AIDS \citep{elliott2015surrogacy}, while in online experimentation platforms diversity of consumed content serves a proxy for long-term visitation
frequencies \citep{wang2022surrogate}. A significant literature exists on designing and analyzing proxy metrics and experiments that use them as a primary outcome. One important question addressed by this literature is choosing (or combining) proxy metrics to be a good surrogate for measuring the effect of the intervention on the long-term outcome \citep{prentice1989surrogate, hohnhold2015focus, parast2017evaluating, NBERw26463, wang2022surrogate, wang2023robust, zhang2023evaluating}. To do so, one needs a principled reason for why the measured treatment effect on the proxy outcome is related to the treatment effect on the long-term outcome. Frequently, this is done by making \emph{causal} assumptions about the relationship between the treatment, proxy outcome, and long-term outcome \citep[see, e.g.,][]{vanderweele2013surrogate,NBERw26463,kallus2022role}. However, motivated by the unique way that trials are run in technology product applications, we take a different approach based on \textit{statistical} regularity assumptions in a population of experiments, similar to meta-analytic approaches such as that taken by \citet{elliott2015surrogacy}.

RCTs performed in technology products are typically referred to as A/B tests. They are used for a wide variety of applications in the technology industry, however one of the most common applications is for assessing the effect of a candidate launch of a new product feature or change on the user's experience. If the results of the A/B test suggest that the candidate launch has a positive effect on the user experience, then it will be deployed to all users. Depending on the scale of the product and engineering team, many candidate launches requiring many A/B tests may be required on a regular basis. The results of these A/B tests on long-term outcomes and proxy metrics are logged, serving as a history of past candidate launches that we may use to guide the choice of future proxy metrics to use for decision making.

This perspective has been studied in the technology research literature previously, for example in \citet{richardson2023pareto, wang2022surrogate, 10.1145/3580305.3599928}, to develop useful heuristics for choosing a proxy metric for use in future A/B tests. In this work, we define a precise statistical framework for choosing a proxy metric based on this historical A/B test data, and develop a method for optimizing a composite proxy, an affine combination of base proxy metrics, that can be used as a primary outcome for future A/B tests. 

The central contributions of our paper are the following:
\begin{itemize}[leftmargin=5mm]
    \item We define a new notion of optimality for a proxy metric we term \textit{proxy quality}, for use in a homogeneous population of randomized experiments. Our definition differs from the existing literature in that it phrases optimality as ensuring the \textit{observed} proxy metric closely tracks the unobserved \textit{population} treatment effect on the long-term outcome (see \cref{fig:proxyquality}). Conveniently, our definition also packages two important considerations for a proxy metric -- short-term sensitivity of the proxy metric and directional alignment with the long-term outcome -- into a single objective (see \cref{eq:single_proxy_quality}). 
    \item We show this new notion of \textit{proxy quality} can be used to construct optimal proxy metrics for new A/B tests via an efficient two-step procedure. The first step reduces the construction of an optimal weighted combination of base proxies, that maximize our definition of proxy quality, to a classic portfolio optimization problem. This optimization problem is a function of the latent variability of the unobservable population treatment effects and the noise level of the new experiment under consideration (see \cref{sec:composite_proxy_quality)}). We then use a hierarchical model to denoise
    the observed treatment effects on the proxy and long-term outcome in a historical corpus of A/B tests
    to extract the variation in the unobserved population treatment effects (see \cref{sec:estimation}). The variance estimates of the population TEs are then used as plug-ins to the aforementioned optimization. 
    \item We highlight the \textit{adaptivity} of our proxy metric procedure to the inherent noise level of each experiment for which it will be used. In our framework the optimal proxy metric for a given experiment is not apriori fixed. Rather it should depend on the sample size (or effective noise level) of the randomized experiment for which it is deployed in order to profitably trade-off bias from disagreement with the long-term outcome and intrinsic variance (see \cref{sec:single_proxy_quality} and \cref{fig:weights_vs_samples}).
    \item Finally, we instantiate and evaluate our framework on a set of 307 real A/B tests from an industrial recommendation engine showing how the proxy metrics we construct can improve decision-making in a large-scale system (see \cref{sec:results}).
\end{itemize}

\subsection{Statistical Framework}
Consider a corpus of $K$ randomized experiments (or A/B tests) where the $i \in \{1, \hdots, K\}$-th experiment is of sample size $n_i$. In each experiment, there is a specific intervention that has some population treatment effect (TE) that we denote $\Delta_i$.\footnote{This may be an average TE (ATE) or relative ATE, but we will not emphasize the differences between these two as they can be handle similarly in our work. For our dataset we use relative ATEs.} In the experiment, we measure an estimated TE $\hat{\Delta}_i$ on the subset of the population included in our experiment. Note that the entire population may be included in the experiment, but $\hat{\Delta}_i$ remains a random quantity, given $\Delta_i$, because of the random assignment of treatments in the experiment.

To differentiate between the TE on the long-term outcome and the proxy metrics, we will attach a superscript $N$ as in $\Delta_i^N$ or $\hat{\Delta}_i^N$ for the long-term outcome (we use $N$ since these are sometimes referred to as north star metrics), and a superscript $P$ as in $\mathbf{\Delta}_i^P$ and $\hat{\bf{\Delta}}_i^P$ for the proxy metrics. Note that there may be multiple base proxy metrics, so $\proxvec{i} \in \mR^d$.
Throughout, we assume that $n_i$ is large enough, and that the experimental design is sufficiently regular such that conditional on $(\Delta_i^N, \proxvec{i})$, $(\hat{\Delta}_i^N, \hatproxvec{i})$ is well-approximated by a Normal distribution centered around the population TE\footnote{Note that in our framework, the notion of proxy quality in \cref{sec:single_proxy_quality} and \cref{sec:composite_proxy_quality)} only relies on low-order moments and doesn't explicitly require Gaussianity (although does use unbiasedness) but the estimation procedure in \cref{sec:estimation} makes explicit use of this structure.} (due to the central limit theorem); and that the joint (within-experiment) covariance of $(\hat{\Delta}_i^N, \hatproxvec{i})$, denoted as $\bm{\Xi}$, has a good estimator, denoted as $\hat{\bm{\Xi}}$.
Our discussion is agnostic to the precise estimator of the TEs and their covariances. We only require black-box access to their values. For the historical corpus of $K$ randomized experiments, we assume that these triplicates of measurements $\{(\hat{\Delta}_i^N, \hat{\bm{\Delta}}_i^P, \hat{\bm{\Xi}}_i)\}_{i=1}^K$ are available. 

Our goal in this paper is to revisit the proxy metric problem: the selection of short-term proxy metrics (or a weighted combination thereof) that track the long-term outcome in a new $K+1$st experiment where measurements of the long-term outcome are unavailable, but measurements of short-term proxy metrics are. In order to develop a statistical framework to construct proxies in a new experiment, we leverage a meta-analytic approach to model the relationship between different experiments. To this end, we assume the population TEs for each experiment are drawn i.i.d. from a common joint distribution $\mathcal{D}$,
\begin{align}
\begin{pmatrix} \Delta^N_i \\ \proxvec{i} \end{pmatrix} & \sim \mathcal{D}(\cdot) \ \text{ i.i.d. } \label{eq:common_joint_dist}
\end{align}
supported over $\mR^{d+1}$. We acknowledge this assumption is strong and not suitable for all applications. However, in our motivating application of interest -- studying a corpus of A/B tests from a large technology company -- historical intuition and various tests do not provide significant evidence this assumption is violated. 
The approach of placing a distributional prior on the population ATE in similar settings of A/B testing at large-scale technology companies \citep{deng2016data, metricsensdecomp} as well as other meta-analytic studies of RCTs \citep{elliott2015surrogacy, elliot2023} has also been advocated for as a useful assumption in prior work.
\section{Methods}
With the above setting in place we first define the measure of quality of a proxy metric which relates the estimated TE on the short-term proxies to the population TE on the long-term outcome for a new experiment. Subsequently, we show how the relevant latent parameters contained in the definition of proxy quality can be efficiently estimated via a hierarchical model.

\subsection{Optimal Proxy Metrics}

\begin{figure}[!htb]
\includegraphics[width=\linewidth, trim={0cm 10cm 0 7cm}, clip]{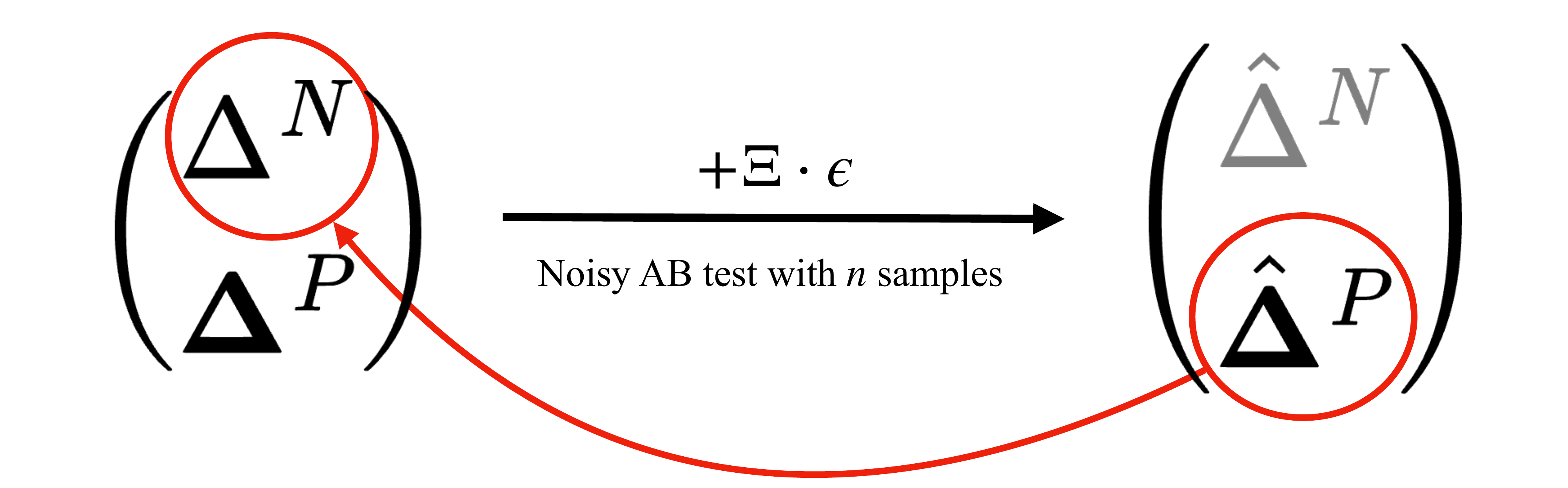}
\caption{In a new experiment, we view the observed TEs as being generated from their corresponding (unobserved) latent values by a noisy channel which adds independent, mean-zero experimental noise with covariance $\bm{\Xi}$. In this new experiment the noisy, observed long-term outcome ${ \hat{\Delta}^N}$ is inaccessible. We seek to find noisy proxy metrics whose TEs closely track the population TEs on the long-term outcome.} 
\label{fig:proxyquality}
\end{figure}

In order to judge the quality of a proxy metric we first define a new notion of utility for a proxy metric. Our key insight is that in a new experiment\footnote{In the following since we assume all the experiments are i.i.d. we suppress index notation on this arbitary experiment drawn from $\cD(\cdot)$.} where $\begin{pmatrix} \Delta^N & \mathbf{\Delta}^P \end{pmatrix}^\top \sim \cD(\cdot)$, the \textit{observed} TEs of the short-term proxies $\hat{\mathbf{\Delta}}^P$ should closely track the \textit{latent} TE of the long-term outcome (see \cref{fig:proxyquality}).\footnote{Note in a new experiment the estimated treatment effect on the long-term outcome $\hat{\Delta}^N$ may be unavailable.}
This is because decisions that are intended to move $\Delta^N$ will be made on the basis of $\hat{\mathbf{\Delta}}^P$.
Thus, we would like these quantities to be well-correlated.

\subsubsection{Proxy Quality of a Single Short-Term Metric in a New Experiment}
\label{sec:single_proxy_quality}
For simplicity, we first consider the case when the vector-valued sequence of proxies reduces to a single scalar proxy. In order to capture the intuition that the short-term estimated proxy TE, $\hat{\Delta}^P$ should track the population long-term outcome TE, $\Delta^N$ we define the \textit{proxy quality} as the correlation between the aforementioned quantities. The correlation is a simple and natural measure which captures the predictive relationship between the proxy metric and long-term outcome. Under stronger conditions in \cref{sec:estimation}, we also argue that optimizing for this measure of proxy quality minimizes the probability of a (signed) decision error or surrogate paradox.

In our setting we consider the case where the estimated TEs are unbiased estimators of their underlying latent population quantities -- so we can parameterize $\hat{\Delta}^P = \Delta^P + \sqrt{\Xi^{PP}} \cdot \epsilon$, where $\epsilon$ is an independent random zero-mean, unit-variance random variable. Hence we can define and simplify the \textit{proxy quality} as:
\begin{align}
    \corr(\Delta^N, \hat{\Delta}^P) & = \frac{\Cov(\Delta^N, \Delta^P)}{\sqrt{\Var(\Delta^N) ( \Var(\Delta^P)+\Xi^{PP})}} 
    \notag \\
    & = \frac{\corr(\Delta^N, \Delta^P)}{\sqrt{1 + \frac{\Xi^{PP}}{\Var(\Delta^P)}}}. \label{eq:single_proxy_quality}
\end{align}
In our setting, the definition of proxy quality decomposes the predictive relationship between the estimated proxy TE and population long-term outcome TE into latent predictive correlation $\corr(\Delta^P, \Delta^N)$ -- a property of the distribution $\cD(\cdot)$ -- and an effective inverse signal-to-noise ratio $\Xi^{PP}/\Var(\Delta^P)$ -- which is also a function of the noise level of the experiment. We now make several comments on the aforementioned quantities.
\begin{itemize}[leftmargin=5mm]
    \item The latent predictive correlation $\corr(\Delta^P, \Delta^N)$ tracks the alignment between the population proxy metric TE and the population long-term outcome TE. In particular, this correlation is reflective of the intrinsic predictive quality of a fixed proxy metric. This quantity is not easily accessible since we do not directly observe data sampled from $\mathcal{D}(\cdot)$. We return to the issue of estimating such quantities in \cref{sec:estimation}.
    \item The quantity $\Xi^{PP}/\Var(\Delta^P)$ computes the ratio of the within-experiment noise in the estimated proxy metric TE---due to fluctuations across experimental units and treatment assignments---to the latent variation of the population proxy metric TE across experiments. For the former quantity we expect $\Xi^{PP}$ to depend on the size of the randomized experiment in consideration (i.e. $\Xi^{PP} \sim \frac{1}{n}$, where $n$ is the sample size of the experiment), since it is a variance over independent treatment units. Meanwhile $\Var(\Delta^P)$ captures how easily the population proxy metric TE moves in the experiment population $\cD(\cdot)$. In a (large enough) given experiment, $\Xi^{PP}$ is easily estimated by $\hat{\Xi}^{PP}$ using the sample covariance estimator.
    Meanwhile, $\Var(\Delta^P)$ is difficult to measure directly, just like $\corr(\Delta^N, \Delta^P)$. We later show how to use a hierarchical model to estimate these parameters (see \cref{sec:estimation}).
    Finally, it is worth noting this ratio term in the denominator is also closely related to a formal definition of metric sensitivity which appears in the A/B testing literature \citep{metricsensdecomp, richardson2023pareto}. 
\end{itemize}
Together the numerator and denominator in \cref{eq:single_proxy_quality} trade off two (often competing) desiderata into a single objective: the numerator favors alignment with the population TE on the long-term outcome while the denominator downweights this by the signal-to-noise ratio of the proxy metric.\footnote{This tradeoff between directional alignment of the proxy metric/long-term outcome and sensitivity of the proxy metric is further discussed in \citep{richardson2023pareto}.} One unique property of this proxy quality measure is that, given a set of base proxies, the ``optimal" single proxy is not an intrinsic property of the proxy metric or distribution of treatment effects captured by $\cD(\cdot)$. Rather, it also depends on the experimental design.  Specifically, it is a function of the experiment sample size $n$, which will control the size of $\Xi^{PP}$. This behavior represents a form of bias-variance trade-off. For large sample sizes, as $\Xi^{PP} \to 0$, \cref{eq:single_proxy_quality} will favor less biased metrics whose population-level TEs are aligned with the long-term outcome (i.e. the numerator is large). Meanwhile, for small sample sizes where $\Xi^{PP}$ is large, \cref{eq:single_proxy_quality} will favor less noisy metrics with a high signal-to-noise ratio so the denominator is small.

\subsubsection{Composite Proxy Quality in a New Experiment}
\label{sec:composite_proxy_quality)}
The previous discussion on assessing the quality of a single proxy metric captures many of the important features behind our approach. However, in practice, we are often not restricted to picking a single proxy metric to approximate the long-term outcome. Rather, we are free to construct a \textbf{composite proxy metric} which is a convex combination of the TEs of a set of base proxy metrics to best predict the effect on the long-term outcome. 

In our framework, the natural extension to the vector-valued setting takes a convex combination of the proxies  $\corr( \Delta^N, \w^\top \mathbf{\hat{\Delta}}^P)$ for a normalized weight vector $\w$, instead of restricting ourselves a single proxy. However, beyond just defining the quality of a weighted proxy metric, we can also optimize for the quality of this weighted sum of base proxy metrics:
\begin{align}
    \max_{\w \in \mR^d} \corr( \Delta^N, \w^\top \mathbf{\hat{\Delta}}^P) : \mathbf{1}^\top \w = 1, \w \geq 0.
\end{align}
Again, the estimated TEs are unbiased estimators of their underlying latent population quantities, so we can parameterize $\mathbf{\hat{\Delta}}^P = \mathbf{\Delta}^P + \left(\bm{\Xi}^{PP}\right)^{1/2} \cdot \bepsilon$, where $\bepsilon$ is an independent random zero-mean, identity-covariance random vector. So the objective expands to:
\begin{align}
    \max_{\w \in \mR^d} \frac{1}{\sqrt{\Var(\Delta^N)}} \frac{\w^\top \Cov(\Delta^N, \mathbf{\Delta}^P)}{ \sqrt{\w^\top(\Cov(\mathbf{\Delta}^P, \mathbf{\Delta}^P)+ \bm{\Xi}^{PP})\w}} : \mathbf{1}^\top \w = 1, \w \geq 0. \label{eq:vector_proxy} 
\end{align}
Essentially all considerations noted in the previous section translate to the vector-valued setting \textit{mutatis mutandis}. In particular, the numerator in \cref{eq:vector_proxy} captures the alignment between the true latent weighted proxy and the population long-term outcome, while the denominator downweights the numerator by the effective noise in each particular experiment. As before we expect $\bm{\Xi}^{PP} \sim \frac{1}{n}$ with the sample size, $n$, of the experiment. Hence, the optimal weights for a given experiment will adapt to the noise level (or equivalently sample size) of the experiment run. 

The formulation in \cref{eq:vector_proxy} raises the question of how to efficiently compute $\w$. Indeed, at first glance the optimization problem as phrased in \cref{eq:vector_proxy} is non-convex. Fortunately, the objective in \cref{eq:vector_proxy} (up to a constant pre-factor) maps exactly onto the Sharpe ratio (or reward-to-volatility ratio) maximization problem often encountered in portfolio optimization \citep{sharpe1966mutual, sharpe1998sharpe}. As is well-known in the optimization literature, the program in \cref{eq:vector_proxy} can be converted to an equivalent (convex) quadratic programming problem which can be efficiently solved \citep[Section 8.2]{cornuejols2006optimization}. We briefly detail this equivalence explicitly in \cref{app:sharpe}. 

The portfolio perspective also lends an additional interpretation to the objective in \cref{eq:vector_proxy}. If we analogize each specific proxy metric as an asset to invest in, then $\Cov(\Delta^N, \mathbf{\Delta}^P)$ is the returns vectors of our assets, and $\Cov(\mathbf{\Delta}^P, \mathbf{\Delta}^P)+ \bm{\Xi}^{PP}$ is their effective covariance--so $\w^\top(\Cov(\mathbf{\Delta}^P, \mathbf{\Delta}^P)+ \bm{\Xi}^{PP})\w$ captures the risk of our portfolio of proxies. Just as in portfolio optimization, where two highly-correlated assets should not be over-invested in, if two proxy metrics are both strongly aligned with the long-term outcome, but are themselves very correlated, the objective in \cref{eq:vector_proxy} will not assign high weights to both of them.

\subsection{Estimation of Latent Parameters via a Hierarchical Model}
\label{sec:estimation}

As the last piece of our framework, we finally turn to the question of obtaining estimates of the unobservable latent quantities arising in \cref{eq:vector_proxy}. While $\bm{\Xi}^{PP}$ is easily estimated from the within-experiment sample covariance $\hat{\bm{\Xi}}^{PP}$, the quantities $\Cov(\Delta^N, \mathbf{\Delta}^P)$, $\Cov(\mathbf{\Delta}^P, \mathbf{\Delta}^P)$ are tied to the latent, unobservable population TEs of the proxy metrics and long-term outcome.

\begin{figure*}[tbh]
\centering     
\subfigure[Latent Population TEs of proxy metric / long-term outcome]{\label{fig:a}\includegraphics[width=0.32 \linewidth]{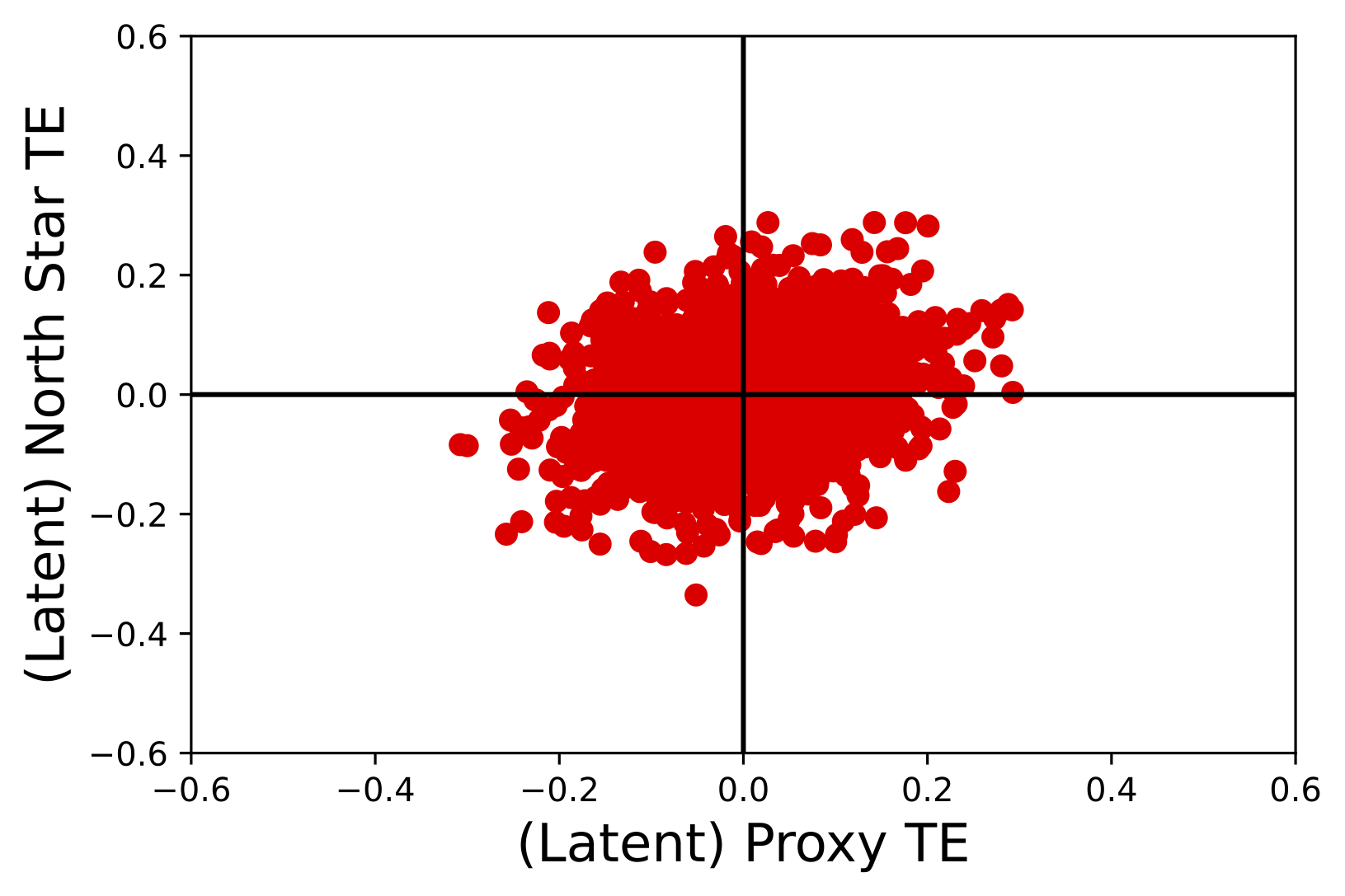}}
\subfigure[Estimated TEs of proxy metric / long-term outcome]{\label{fig:b}\includegraphics[width=0.32 \linewidth]{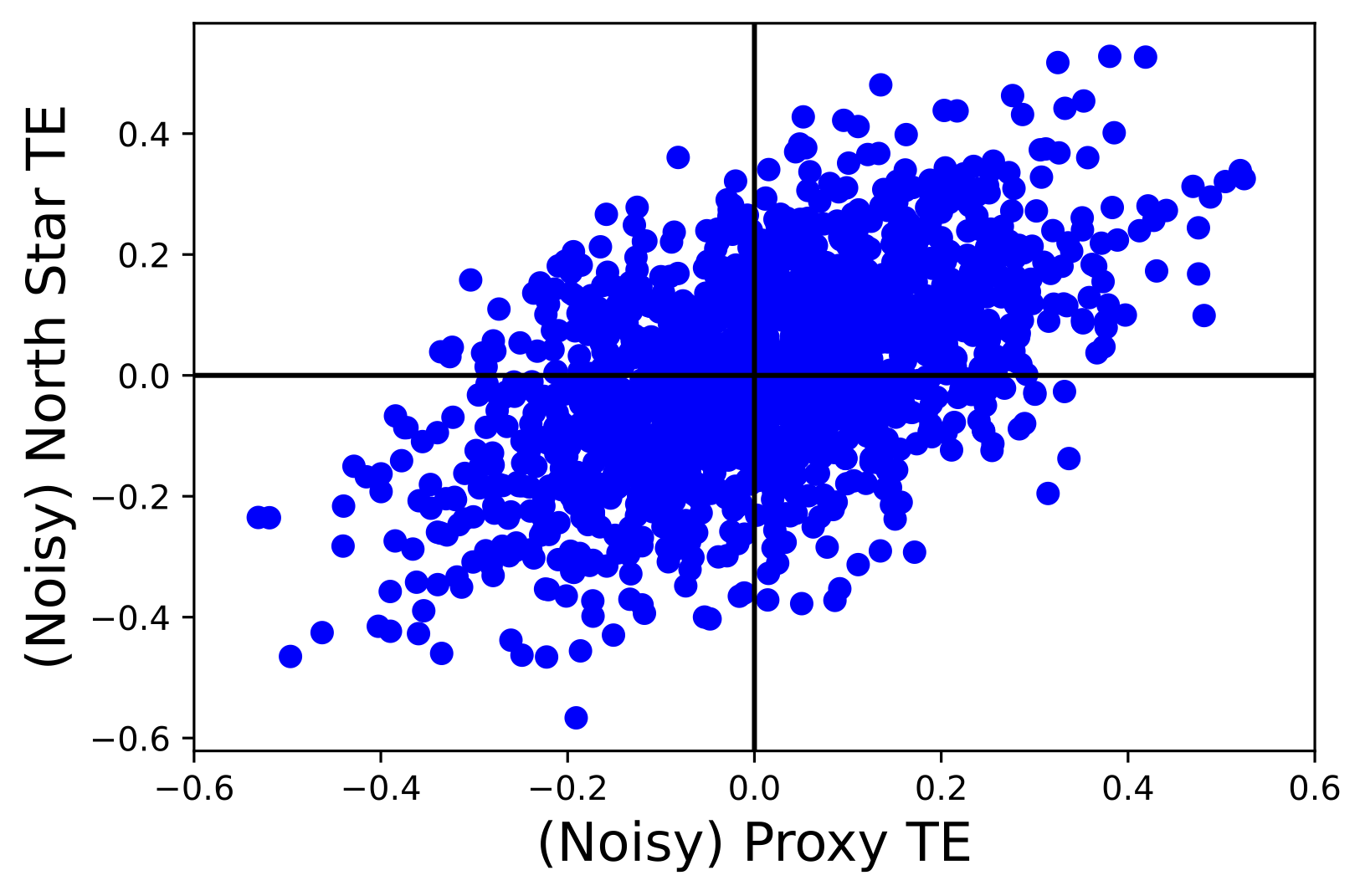}}
\subfigure[Estimated TEs with raw / estimated covariance from hierarchical model]{\label{fig:c}\includegraphics[width=0.32 \linewidth]{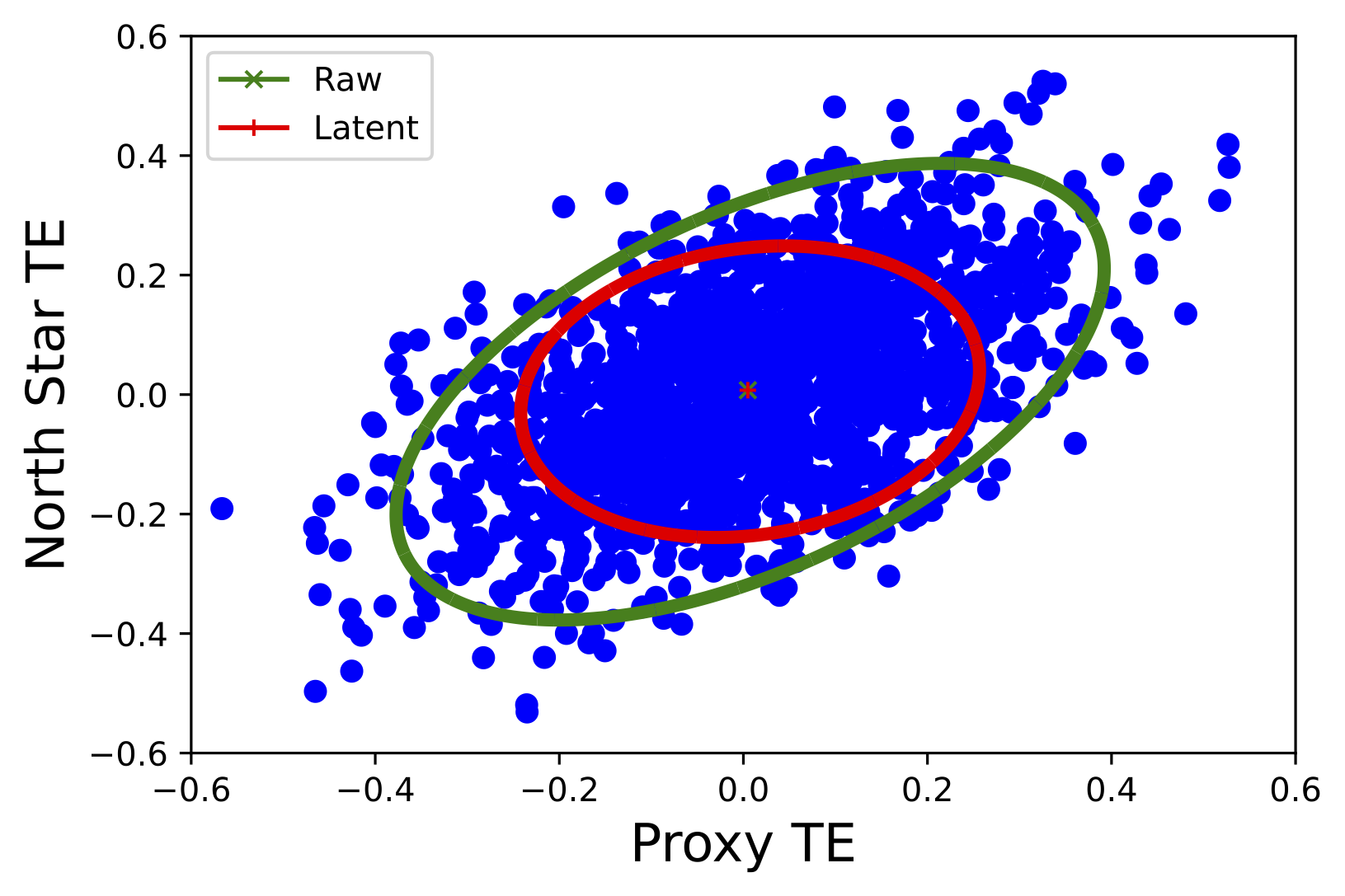}}
\caption{The panel visualizes the denoising effect of fitting a hierarchical model to raw TEs to uncover their latent variation on synthetic data. We generate 1500 synthetic datapoints sampled from the model in \cref{eq:level1} with one proxy metric. Each datapoint represents a synthetic TE measurement from a single A/B test. We use parameters with $\begin{pmatrix}\mu^N \\ \mu^P \end{pmatrix} = \begin{pmatrix} 0.0 \\ 0.0 \end{pmatrix}$,  $\bm{\Lambda} = .01 \cdot \begin{pmatrix} 1 & 0.2 \\ 0.2 & 1 \end{pmatrix}$ to generate data in \cref{fig:a}. We add Gaussian noise with covariance $\bm{\Xi} = .02 \cdot \begin{pmatrix} 1 & 0.7 \\ 0.7 & 1 \end{pmatrix}$ to them in \cref{fig:b}. Finally, we fit the generative model to the observed data in \cref{eq:hm} using the within-experiment covariances $\bm{\Xi}$ in \cref{fig:c}. \cref{fig:c} illustrates how the hierarchical model denoises the raw observed TEs to disentangle the latent variation in the population from the experimental noise in each synthetic A/B test.}
\label{fig:denoising}
\end{figure*}

In order to gain a handle on these quantities, we take a meta-analytic approach which combines two key pieces. First, as in our previous discussion, we require the setting described in \cref{eq:common_joint_dist} -- that is we assume the true population TEs are drawn i.i.d. from a common joint distribution. While only this assumption was needed for our previous discussion, we now introduce additional parametric structure in the form of an explicit generative model to allow for tractable estimation of the parameters $\Cov(\Delta^N, \mathbf{\Delta}^P)$ and $\Cov(\mathbf{\Delta}^P, \mathbf{\Delta}^P)$. Second, we assume access to a pool of homogeneous RCTs for which unbiased estimates of the TE on the short-term proxy metrics and long-term outcome are available (i.e. $(\hat \Delta^N_i, \hatproxvec{i})$ for $i \in \{1, \hdots, K \}$). With these two pieces we can construct a hierarchical (or linear mixed) model to estimate the latent parameters:
\begin{align}
\begin{pmatrix} \Delta^N_i \\ \mathbf{\Delta}^P_i\end{pmatrix} &\sim
\mathrm{MVN} \left(\begin{pmatrix} \mu^N \\ \bm{\mu}^P \end{pmatrix}, \bm{\Lambda} \right) \label{eq:level1} \\
\begin{pmatrix} \hat \Delta^N_i \\ \hat{\mathbf{\Delta}}^P_i \end{pmatrix} \mid \begin{pmatrix}\Delta^N_i \\ \mathbf{\Delta}^P_i \end{pmatrix} &\sim
\mathrm{MVN}\left(\begin{pmatrix} \Delta^N_i \\ \mathbf{\Delta}^P_i \end{pmatrix}, \bm{\Xi}_i\right),  \quad \forall i \in [K]. \label{eq:level2}
\end{align}
which due to the joint Gaussianity we can marginalize as:
\begin{equation}
\begin{pmatrix}\hat \Delta^N_i \\ \hat{\mathbf{\Delta}}^P_i\end{pmatrix} \sim
\mathrm{MVN}\left(\begin{pmatrix}\mu^N \\ \bm{\mu}^P\end{pmatrix}, \bm{\Sigma}_i := \bm{\Lambda} +  \bm{\Xi}_i\right), \quad \forall i \in [K]. \label{eq:hm}
\end{equation}
We use the notation $\bm{\Lambda}$ to capture the latent covariance of the joint distribution $\cD(\cdot)$ which we parameterize by a multivariate normal (i.e. MVN). So in our case, $\Cov(\Delta^N, \mathbf{\Delta}^P) = \bm{\Lambda}^{NP}$ and $\Cov(\mathbf{\Delta}^P, \mathbf{\Delta}^P) =\bm{\Lambda}^{PP}$ for $\begin{pmatrix} \Delta^N & \mathbf{\Delta}^P \end{pmatrix}^\top \sim \cD(\cdot)$. Moreover, for the purposes of inference we simply use the plug-in estimate $\hat{\bm{\Xi}} \approx \bm{\Xi}$ which is routinely done in similar hierarchical modeling approaches \citep{gelman1995bayesian}. While using multivariate normality in \cref{eq:level1} is an assumption (albeit we believe reasonable in our case), it is not essential to the content of our results. Our proxy quality definition relies only on inferring low-order moments of $\cD(\cdot)$ for which this parametric structure is convenient. The second approximation that the noisy TEs are multivariate normal around their true latent values (\cref{eq:level2}) is well-justified by the central limit theorem in our case, since the experiments we consider all have at least $O(10^5)$ treatment units. Since inference in this model is not closed-form, we implement the aforementioned generative model in the open-source probabilistic programming language NumPyro \citep{phan2019composable} to extract the latent parameters\footnote{As an alternative to method presented, we could eschew the parametric Gaussian structure by using an estimator which uses sample-splitting \textit{within each RCT} to estimate $\bm{\Lambda}$. We prefer to use our current approach in order to implicitly reweight by the heteroscedasticity in our observations (i.e. $\Xi_i$) and avoid sample-splitting/cross-fitting. Additional details are provide in \cref{app:lmm}.}. Additional details on the inference procedure are deferred to \cref{app:lmm}.

\cref{fig:denoising} provides an example where the inference procedure is used to extract the latent population variation in a synthetically generated dataset. Although this example is synthetic (and exaggerated), empirically in our corpus we observe many base proxy metrics with correlations to the long-term outcome of $\sim 0.6$ in their experimental noise matrix $\hat{\bm{\Xi}}$. Thus, the example shows a case where the raw correlation may provide an over-optimistic estimate of the underlying alignment between a proxy metric and long-term outcome. The denoising model we fit helps mitigate the impact of correlated within-experiment noise in our setting. We schematically detail the end-to-end algorithm which composes the denoising model fit and portfolio optimization to construct a proxy for a new A/B test in \cref{algo:composite_index}.

Lastly, with the generative model in \cref{eq:level1,,eq:level2} in place we can provide an alternative interpretation of our definition of composite proxy quality. Under the condition that $\begin{pmatrix}\mu^N \\ \bm{\mu}^P \end{pmatrix} = \mathbf{0}$\footnote{This condition may not be true in all applications but is approximately satisfied in our dataset with all metrics having the ratio of their global mean to global standard deviation being bounded by $0.1$ but often being even less.}, the probability of a signed alignment (or equivalently the complement of a surrogate paradox \citet{elliott2015surrogacy}) can be simplified too,
\begin{align}
    \Pr (\Delta^N > 0,\w^\top \mathbf{\hat \Delta}^P > 0 ) =  \Pr(\Delta^N < 0,\w^\top \mathbf{\hat \Delta}^P <0) = \frac{1}{4} + \frac{\sin^{-1}(\rho)}{2 \pi}
\end{align}
for $\rho = \frac{1}{\sqrt{\Lambda^{NN}}}\frac{\w^\top \bm{\Lambda}^{N P}}{\sqrt{\w^\top ( \bm{\Lambda}^{PP}+\bm{\Xi}^{PP}) \w}}$. This computation relies on the joint Gaussianity of the model and centering condition to explicitly compute the alignment probability. Given $\rho$ is our definition of the vector-valued proxy quality and the inverse-sine function is monotone increasing, optimizing the proxy quality over $\w$ can be interpreted as minimizing the probability of a signed decision error in this setting.

\begin{algorithm}[!bt]
\caption{Composite Proxy Algorithm}\label{algo:composite_index}
\begin{algorithmic}[1]
\renewcommand{\algorithmicrequire}{\textbf{Input: }}
\renewcommand{\algorithmicensure}{\textbf{Output: }}
\REQUIRE $\{(\hat{\Delta}_i^N, \hat{\bm{\Delta}}_i^P, \hat{\bm{\Xi}}_i)\}_{i=1}^K$ (TE and Noise Estimates from Historical Tests),
$\hat{\bm{\Xi}}_{K+1}^{PP}$ (Noise Estimate for New Test).

\STATE $\bm{\mu}, \bm{\Lambda} \leftarrow$ \textsc{HM}($\{(\hat{\Delta}_i^N, \hat{\bm{\Delta}}_i^P, \hat{\bm{\Xi}}_i)\}_{i=1}^K$). Fit model (e.g. using Numpyro) defined in \cref{eq:hm}.
\STATE $\w \leftarrow$ \textsc{Sharpe Ratio}($\bm{\Lambda}, \hat{\bm{\Xi}}_{K+1}^{PP}$). Solve convex program in \cref{eq:qp} (e.g. using CVXPY).
\ENSURE $\w$, $\w^\top \hat{\bm{\Delta}}_{K+1}^P$ (Proxy Weights and Composite Proxy for New Experiment).
\end{algorithmic}
\end{algorithm} 

\section{Results}
\label{sec:results}
In this section, we turn to evaluating the performance of our composite proxy procedure against several baselines. As raw proxy metrics to consider in our evaluations we use a small set of 3 hand-selected proxy metrics which capture different properties domain experts believe are relevant to long-term user satisfaction in our setting. We first highlight a unique feature of our proxy procedure -- its adaptivity to the sample size of the experiment for which it will be applied. We then perform a comparison of our new proxy procedure against the raw proxy metrics and a baseline procedure \citep{richardson2023pareto} appearing in the literature. 

\subsection{Proxy Quality and Sample Size Dependence}
\label{sec:sample_size}

\begin{figure*}[tbh]
\centering
\includegraphics[width=0.8\linewidth, trim={0cm 0cm 0 0cm}, clip]{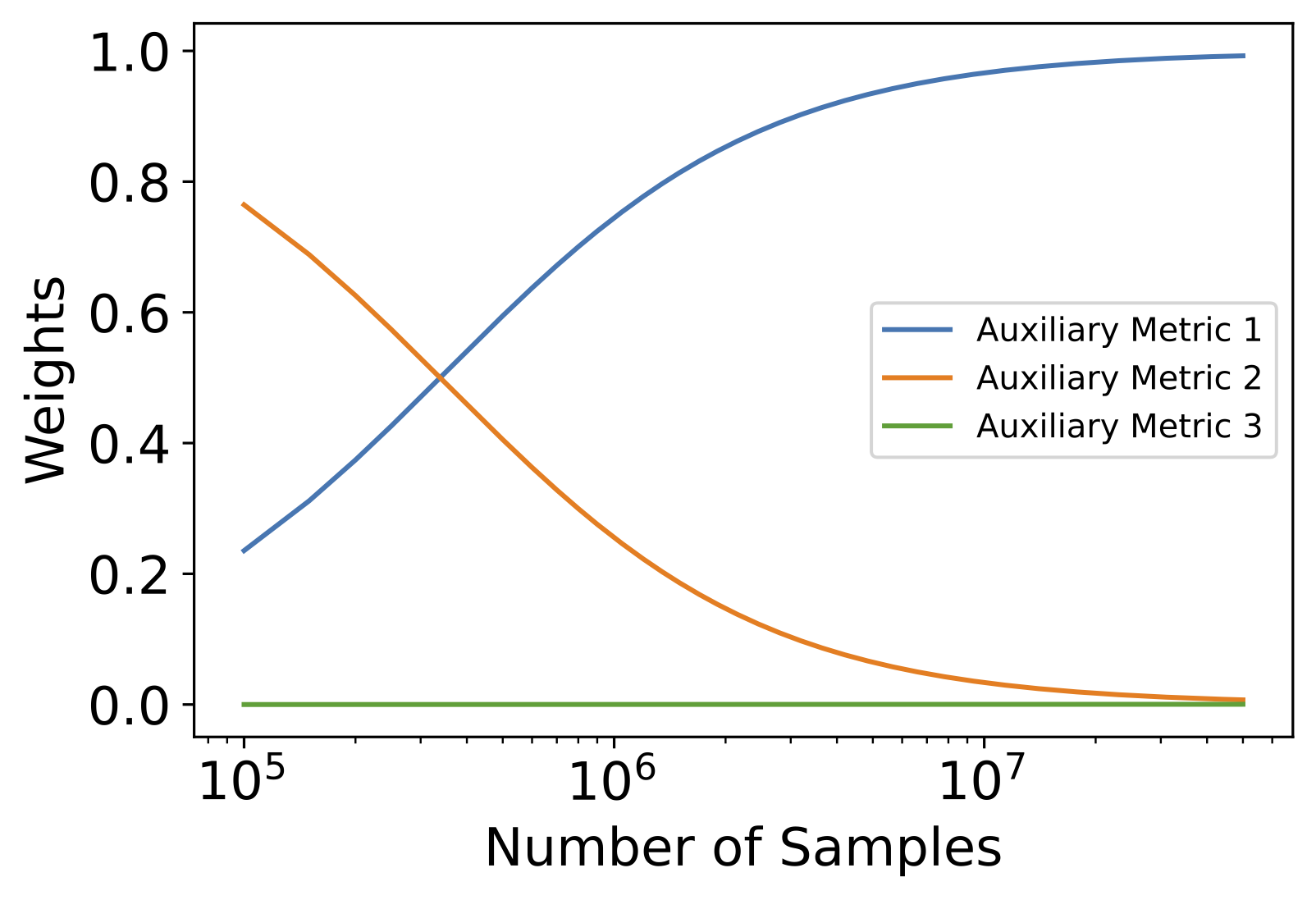}
\caption{The optimal weighting dependence on sample size for our new composite proxy, represents a bias-variance trade-off. For large sample sizes the weighting favors potentially noisier metrics that are more aligned with the long term outcome. However, for smaller sample sizes the optimal weighting backs off to metrics which are less noisy but also less aligned to the long term outcome.} 
\label{fig:weights_vs_samples}
\end{figure*}

One unique feature of our procedure is its adaptivity to the noise level (or effectively sample size of the experiment); recall in \cref{eq:vector_proxy} the optimal weights will depend on the latent parameters which are inferred from the pool of homogeneous RCTs on which they are fit, but also the experiment noise estimate $\hat{\bm{\Xi}}$ which depends on the new A/B test it is to be used for. While in practice one could recompute a proxy metric depending on the \textit{aposteriori} results of each A/B test (so $\hat{\bm{\Xi}}$ is known), it is also desirable to be able to fit a proxy for each A/B test \textit{apriori}, without knowledge of its results. To do so, we found that in our application, $\bm{\Xi}^{PP}_i$ could be estimated with reasonable accuracy purely on the basis of historical data of \textit{other A/B tests} in our population of experiments by postulating a scaling of the form $\bm{\Xi}^{PP}_i = \bm{\Xi}^{PP}_{\rref}/n_i$. Here the reference matrix $\bm{\Xi}^{PP}_{\rref}$ can be thought of as the within experiment variance of an A/B test in the population with one sample. The ansatz $\bm{\Xi}^{PP}_i = \bm{\Xi}^{PP}_{\rref}/n_i$, combines two observations. The first is that the variance of a TE estimate decays as $\sim \frac{1}{n_i}$ in the number of treatment units $n_i$, which is immediate from the independence of treatment units. However, the second is that the constant prefactor in the variance $\bm{\Xi}^{PP}_{\rref}$ is approximately the same across different A/B tests in our corpus. The reference matrix $\bm{\Xi}^{PP}_{\rref}$ can then be estimated as a weighted average of $\hat{\bm{\Xi}}_i$ from the corpus. Additional details and verification of these hypotheses are provided in \cref{app:scaling}. The upshot of this approach is that the computation of the optimal weights in \cref{eq:vector_proxy} for a new A/B test can then be done using only the sample size ($n_i$) of this new experiment (i.e. before the new experiment is ``run").

To understand the dependence of our new composite proxies weighting on the experiment sample size, we fit the latent parameters $\bm{\Lambda}$ (from the hierarchical model in \cref{eq:level1,,eq:level2}) and $\bm{\Xi}^{PP}_{\rref}$ on the entire corpus for the results in \cref{fig:weights_vs_samples}. We then use the scaling $\bm{\Xi} = \bm{\Xi}^{PP}_{\rref}/n$ to estimate the optimal weights of our new composite proxy from \cref{eq:vector_proxy} for different sample sizes $n$ for a hypothetic new A/B test. \cref{fig:weights_vs_samples} shows how as the sample size increases the new composite proxy smoothly increases its weighting on raw metrics which are noisier but more strongly correlated with the long term outcome. Moreover, while the Auxiliary Metric 3 is an intuitively reasonable metric, its value as determined by the measure of proxy quality is dominated by a mixture of the other two components.

\subsection{Held-out Evaluation of Proxy Procedures}
\label{sec:eval}

\begin{table*}[htb]
\centering
\begin{tabular}{llrrr}
\toprule
                            Metric &  Sensitivity &  Proxy Score &  Proxy Quality \\
\midrule
                New Composite Proxy &     0.181 &     0.666 &       0.302 \\
          Baseline Composite Proxy &     0.182 &     0.611 &       0.279 \\
        Auxiliary Metric 1 &     0.062 &     0.611 &       0.174 \\
            Auxiliary Metric 2 &     0.368 &     0.222 &       0.258 \\
  Auxiliary Metric 3 &     0.166 &     0.104 &       0.030 \\
\bottomrule
\end{tabular}
\caption{Comparison of our new composite proxy against a baseline method, and their constituent base proxies alongst several criterion which are computed on a held-out evaluation. Our new composite proxy performs favorably across all measures -- notably achieving the highest proxy score and proxy quality amongst all considered. All evaluation criteria are bounded in $[0,1]$ and for each higher is better.}
\label{table:comparison}
\end{table*}

The primary difficulty of this evaluation is that in TE estimation there is the lack of ground-truth ``labels" of the treatment effect (i.e. in our framework the population latent TEs such as $\Delta^N$ are never observed). However, in our setting we do have access to a large corpus of 307 A/B tests as noted earlier. Hence, we use held-out/cross-validated evaluations of certain criterion which depend on the noisy metrics aggregated over an evaluation set, to gauge the performance of proxy metrics fit on a training set. 

We consider several relevant criteria for performance which have been used in the literature. Two important measures which appear in \citep{richardson2023pareto} are the proxy score and sensitivity. To define the criteria, recall that a TE metric is often used to make a downstream decision by thresholding its t-statistic, $\tstat = \frac{\hat{\Delta}}{\sqrt{\Var(\hat{\Delta})}}$ as $\tstat > 2 \to + \text{ (positive)}$, $-2 < \tstat < 2 \to 0 \text{ (neutral)}$, and $\tstat < -2 \to - \text{ (negative)}$. Given a corpus of A/B tests we can then compute the decisions induced by a short-term proxy metric, the decisions induced by the long-term outcome and check the number of A/B tests for which they align. After normalization, the number of detections (both metrics decisions are positive or negative) minus the number of mistakes (one metric decision is positive while the other is negative) defines the proxy score. Similarly, for a short-term proxy metric we can compute its sensitivity -- which is the number of times it triggers a statistically significant decision by being positive or negative. Loosely speaking, these two criterion function like the notions of precision and recall in information retrieval. Ideally, a short-term metric would maximize both quantities by being sensitive and triggering often (so as to not miss any A/B tests where the TE for the long-term outcome is significant) but not over triggering and leading to many false positives (or negatives). Additional details on these metrics are provided in the \cref{app:proxy_score}\footnote{It is worth noting these measures are still imperfect in the sense that comparisons are made against the decisions induced by the noisy estimate of the long-term outcome TE not the latent population long-term outcome TE. 
}. As another measure of performance, we also compute and report our definition of our composite proxy quality for a given composite proxy's learned weights.


We use the same set of raw proxy metrics as before in our evaluation. As a procedure to compare our methodology against, we use the baseline of optimizing the convex combination of these 3 metrics to optimize the aforementioned proxy score which is detailed in \citep{richardson2023pareto}. In each case we use stratified 4-fold cross-validation (CV) to compute the weights for each procedure on a training subset of the corpus and evaluate the metrics on the held-out set by computing the aforementioned evaluation scores. The baseline proxy method and base proxy metrics each learn a fixed set of weights\footnote{the base proxy metrics simply place all their weight on themselves.} depending on the training fold, which is applied to each A/B test in the evaluation fold. For our procedure we use the ansatz $\bm{\Xi}^{PP}_i = \bm{\Xi}^{PP}_{\rref}/n_i$ mentioned in the previous setting to calculate $\bm{\Xi}^{PP}_{\rref}$ as a simple weighted average from data only in the training fold of our CV split. The optimal proxy metric for each A/B test in the test fold can then be refit using only the sample size $n_i$ of that A/B test. This strategy has the additional benefit of enforcing strict separation of the data in train/test set folds in our CV split.

Results for our cross-validated evaluation are displayed in \cref{table:comparison} across our corpus of 307 historical A/B tests which come from a real industrial recommendation engine. Note that both our new composite proxy and the baseline composite proxy improve significantly in the proxy score and proxy quality over the raw metrics without sacrificing unduly on sensitivity relative to Auxiliary Metric 2. Moreover, the new composite proxy achieves not only the highest proxy quality but also proxy score despite not explicitly optimizing for proxy score on the training set. We believe this may be a feature of our new composite proxy  whose weights are adaptive to the size of each experiment in the A/B test (which vary in our corpus from approximately $O(10^6)$ to $O(10^8)$ in size).

\section{Conclusion}
We have presented a framework for both defining and constructing optimal composite proxy metrics, which are used to approximate the decisions induced by a difficult-to-measure long-term metric. One key insight from our framework is that the optimal proxy for a given experiment should depend on the noise level (or equivalently the size) of that experiment. In our work, the first component of our procedure reduces the composite proxy selection problem to a portfolio optimization relating the unobserved long-term/north star TE and observed TE. This does not explicitly use the population distribution assumption in \cref{eq:common_joint_dist}, although it relies on the unobservable treatment effects, and we believe this is a valuable and natural framing of the proxy selection problem. \cref{eq:common_joint_dist} is implicitly needed to identify the latent covariance parameters between a new future RCT and past RCTs, and accordingly for estimation of the unobserved latents from the corpus of RCTs. Due to the lack of ground truth (true treatment effects are unobserved in RCTs) \cref{eq:common_joint_dist} is important for our meta-analysis, although we acknowledge that it is not suitable for all applications. One of the limitations of our work is that highly non-stationary settings (where \cref{eq:common_joint_dist} is not a good approximation) may not be well addressed by our second-stage denoising procedure. Accordingly, this assumption should be probed by model-fit diagnostics and domain-specific intuition. 

An interesting direction for future work is to further relax this assumption through different structural or causal assumptions which are application-dependent in the second part of the procedure (while still maintaining the first portfolio optimization component). Here, more flexible modeling of the joint latent effect distribution -- to accommodate more structured latent effects could be useful. Extending our framework to handle the construction of nonlinear composite proxy metrics in situations where higher-order interactions between the north star and short-term proxies are important, is also an interesting avenue for further research. As an additional point of exploration, understanding how our approach might generalize to the problem of heterogeneous treatment effect estimation -- to provide contextual decision-making power -- would also be valuable.
\section{Acknowledgements}
We thank Jasper Snoek and our anonymous reviewers for their feedback on this paper.

\appendix
\newpage
\section{Within-Experiment Covariance Scaling}
\label{app:scaling}
As we note before, one interesting feature of our composite proxy quality procedure is its dependence on the noise level of $\bm{\Xi}$ of the randomized experiment it will be applied too. Recall in \cref{eq:vector_proxy} the optimal weights will depend not only on the latent parameters which are instrinsic properties of $\cD(\cdot)$, but the experiment noise estimate $\hat{\bm{\Xi}}$ which depends on the particular A/B test too which it is applied.

\begin{figure}[!htb]
\centering     
\subfigure[Variance of A/B tests for Auxiliary Metric 1]{\label{fig:a2}\includegraphics[width=0.48 \linewidth]{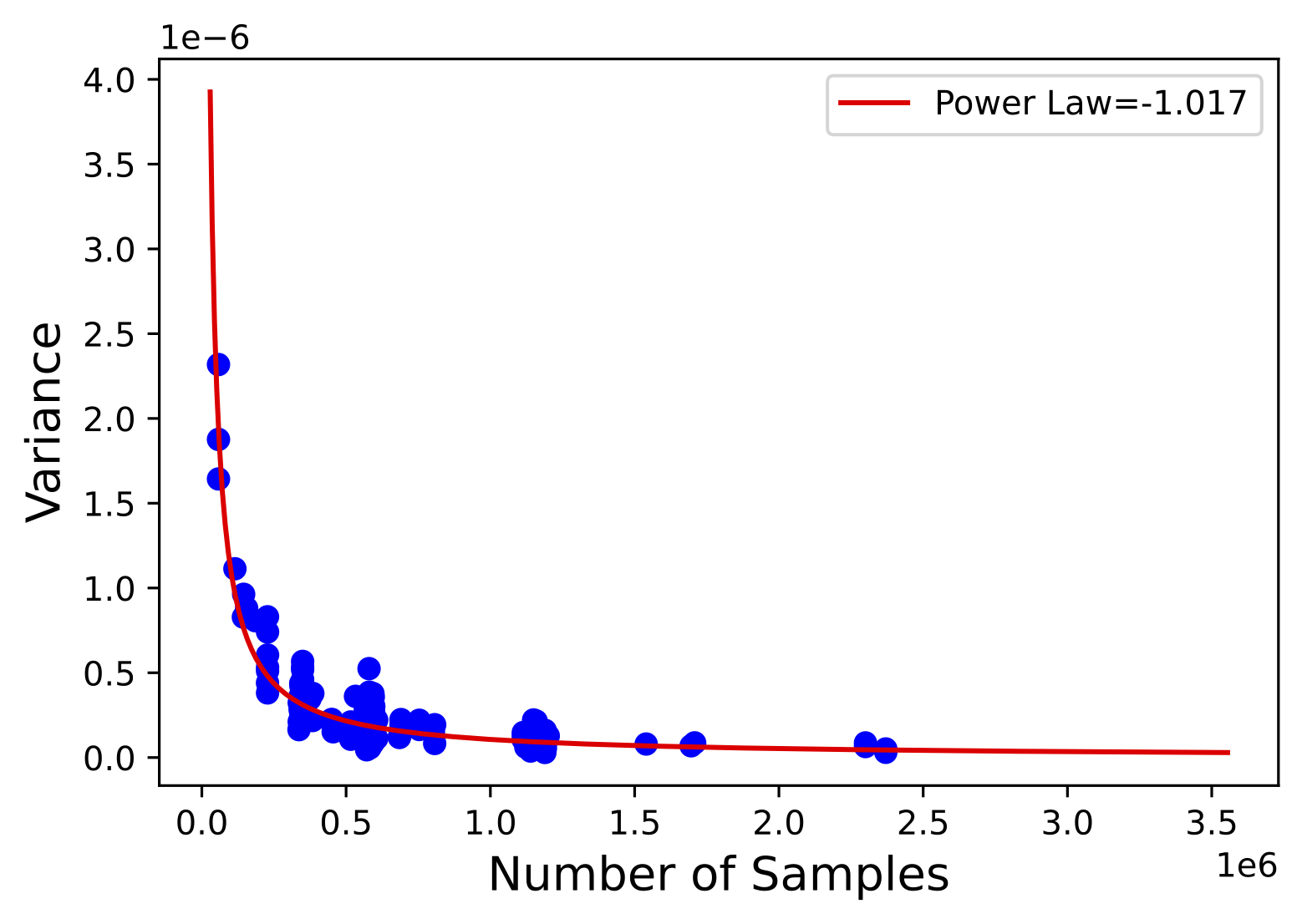}}
\subfigure[Variance of A/B tests for Auxiliary Metric 2]{\label{fig:b2}\includegraphics[width=0.48 \linewidth]{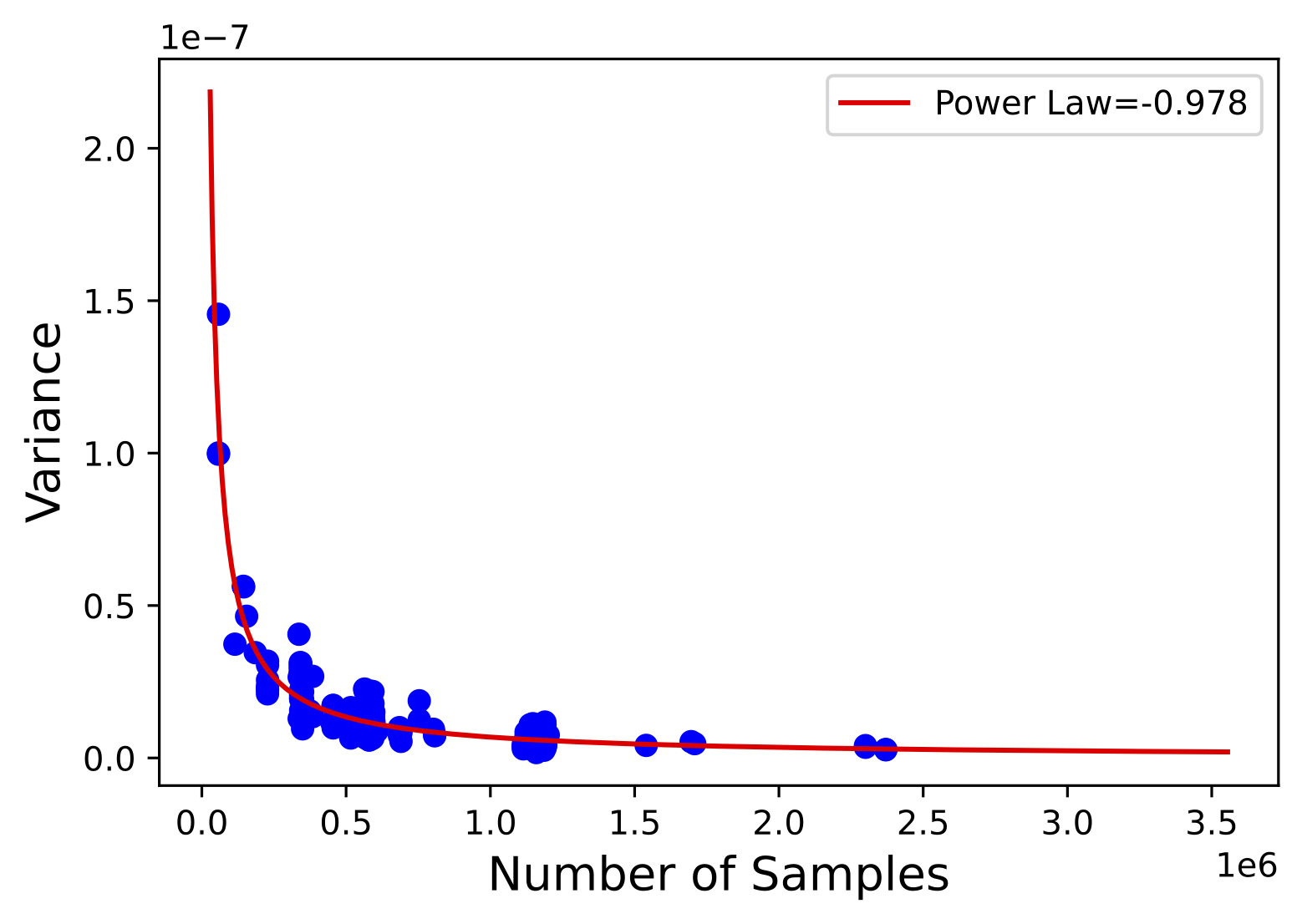}}
\caption{Both displays show the within-experiment marginal sample variance (blue dots) for two different metrics computed across 307 different A/B tests and their corresponding power-law fit (red line). Despite the underlying A/B tests being different, we found that the variance were reasonably well modeled by a single inverse-power law with the same constant prefactor over the entire population.}
\label{fig:var_scaling}
\end{figure}

While in practice, one could recompute a composite proxy metric after an A/B test is run (so $\hat{\bm{\Xi}}$ is known), in many applications it is also desirable to be able to fit weights for a composite proxy metric before each A/B test is run. In order to do this we found that we could build a simple predictive model for the experimental noise level $\bm{\Xi}^{PP}_i$ in a given experiment on the basis of historical data of \textit{other A/B tests} in our population of experiments. We did so by making an ansatz of the form $\bm{\Xi}^{PP}_i = \bm{\Xi}^{PP}_{\rref}/n_i$, where $\bm{\Xi}^{PP}_{\rref}$ can be thought of as the within experiment variance of an A/B test in the population with one sample. This ansatz follows from two facts. The first is that the variance of a TE estimate decays as $\sim \frac{1}{n_i}$ in the number of treatment units $n_i$, which is immediate from the independence of treatment units. However, the second -- that the constant prefactor in the variance $\bm{\Xi}^{PP}_{\rref}$ is approximately the same across different A/B tests -- is an empirical observation due to underlying homogeneity in the population of A/B tests. This approximate homogeneity is evidenced in \cref{fig:var_scaling}.

On the basis of this ansatz for each A/B test,
\begin{align*}
    \bm{\Xi}_i = \frac{\bm{\Xi}_{\rref}}{n_i}, \quad \{1, \hdots, K \}
\end{align*}
we can construct an estimator for the reference constant matrix $\bm{\Xi}_{\rref}$ as,
\begin{align}
    \hat{\bm{\Xi}}_{\rref} = \sum_{i=1}^{K} \gamma_i n_i \hat{\bm{\Xi}}_i
\end{align}
for some convex combination of weights $\gamma_i$. While we can use an equal weighting scheme where $\gamma_i = \frac{1}{K}$, since the sample variance estimates $\hat{\bm{\Xi}}_i$ themselves are noisy, we instead use a precision weighted combination of them to reduce variance by taking $\gamma_i \propto n_i$. 

With this estimate $\hat{\bm{\Xi}}_{\rref}$ in hand, for a new $K+1$st A/B test with sample size $n_{K+1}$ we can approximates its within-experiment covariance as,
\begin{align*}
    \hat{\bm{\Xi}}_{K+1} \approx \frac{\hat{\bm{\Xi}}_{\rref}}{n_{K+1}}
\end{align*}
under the implicit homogeneity assumption we make.

\section{Sharpe Ratio and Portfolio Optimization}
\label{app:sharpe}

Under the mild condition (which is always satisified for us that at least one element of $\bm{\Lambda}^{NP}$ is positive),
we can transform the Sharpe ratio optimization objective to an equivalent convex quadratic program:
\begin{align}
    \min_{\x \in \mR^d} \x^\top \bm{\Sigma} \x \quad : \quad \x \geq 0, \quad \mathbf{r}^\top \x = 1. \label{eq:qp}
\end{align}
where $\bm{\Sigma} = \bm{\Lambda}^{PP} + \hat{\bm{\Xi}}^{PP}$ and $\mathbf{r}= \bm{\Lambda}^{NP}$. The solution to the original problem in \cref{eq:vector_proxy} can be recovered by normalizing as $\w = \frac{\x}{\norm{\x}_1}$. The details of this standard transformation can be found in \citep[Section 8.2]{cornuejols2006optimization}, although the original reduction is a generalization of the Charnes-Cooper transformation \citep{charnes1962programming} which dates back to at least \citep{schaible1974parameter}.

\section{Inference in Hierarchical Model}
\label{app:lmm}
In order to extract estimates of the latent parameter we perform full Bayesian inference over the generative model in \cref{eq:hm} using Numpyro \citet{phan2019composable} which uses the NUTS sampler to perform MCMC on the posterior. We found Bayesian inference to be more stable then estimating the MLE of the model. We augmented the generative model in  \cref{eq:hm} with the weak priors:
\begin{align*}
    & \bm{\mu} \sim \cN(0, 1000 \cdot \text{meanscale} \cdot  \I_{d+1}) \\
    & \mathbf{s} \sim 1.5 \cdot \textsc{Half-Cauchy}(\text{devscale}) \\
    & \mathbf{C} \sim \textsc{LKJ}(\text{concentration}=1) \\
    & \bm{\Lambda} = \sqrt{\mathbf{s}}^\top \circ \mathbf{C} \circ \sqrt{\mathbf{s}}
\end{align*}
where we use the operator $\circ$ to denote coordinatewise broadcasted multiplication. Here the vector-valued parameters \text{meanscale}, and \text{devscale} are set to match the overall scales of the raw mean and raw covariance of the corpous A/B tests. We found the overall inferences to be robust to the choice of scales in the Half-Cauchy prior on the pooled variance parameter and normal prior mean, which are both weakly-informative.  \citep{gelman2006prior} and \citep{polson2012half} both argue for the use of the Half-Cauchy prior for the top-level scale parameter in hierarchical linear models as opposed to the more traditional use of the Inverse-Wishart prior on both empirical and theoretical grounds. The choice of the LKJ prior with concentration parameter set to 1 is essentially a uniform prior over the space of correlation matrices \citep{gelman1995bayesian, lewandowski2009generating}.

Inference in this model was performed using the default configuration of the NUTS sampler in Numpyro \citep{phan2019composable}. We also found it useful to initialize the parameters $(\mu^N, \bm{\mu}^P)$ and $\bm{\Lambda}$ to the scales of the raw mean and raw covariance of the corpus A/B tests. We diagnosed convergence and mixing of the NUTS sampler using standard diagnostics such as the r-hat statistic \citep{gelman1995bayesian}. In all our experiments we found the sampler mixed efficiently and we achieved a perfect r-hat statistic for all parameters of 1.0. For each MCMC run we generated 10000 burn-in samples and 50000 MCMC samples for 4 parallel chains. We used the posterior means of the samples to extract estimates of $\bm{\Lambda}$ for use in our proxy quality score.

As noted in the main text an alternative to using the generative model presented here which eschews the Gaussian parametric structure (but of course relies on the \cref{eq:common_joint_dist}) is to use a sample-splitting estimator \textit{within each RCT}. Our given procedure is agnostic to the details of TE estimation so long as each estimate is unbiased. However for the present discussion, assume as before, we have a corpus of RCTs with true TEs satisfying \cref{eq:common_joint_dist}, 
\begin{align}
\mathbf{\Delta}_i = \begin{pmatrix} \Delta^N_i \\ \proxvec{i} \end{pmatrix} & \sim \mathcal{D}(\cdot) \ \text{ i.i.d.}, i \in \{1, \hdots, n \}, \notag
\end{align}
with mean and covariance vectors $\bm{\mu}$ and $\bm{\Lambda}$. Further assume in each RCT we have two unbiased estimates for the TE satisfying $\mathbf{\hat{\Delta}}_{i,1} = \mathbf{\Delta}_i + \bm{\Xi}^{1/2} \cdot \bepsilon_{i,1}$ and $\mathbf{\hat{\Delta}}_{i,2} = \mathbf{\Delta}_i + \bm{\Xi}^{1/2} \cdot \bepsilon_{i,2}$ with mutually independent mean-zero observation noise $\bepsilon_{i,1}, \bepsilon_{i,2}$. Such estimators can easily be obtained by randomly splitting the units in treatment and control groups in the RCT into two disjoint subsets and computing an unbiased TE estimate (such as the difference-of-means estimator) on each subset. Since they are from same RCT they will provide unbiased estimates of the same unobserved TE $\bm{\Delta}_i$. So it follows that $\mE[\mathbf{\hat{\Delta}}_{i,1} \mathbf{\hat{\Delta}}_{i,2}^\top] = \bm{\Lambda} + \bm{\mu} \bm{\mu}^\top$ for each $i \in \{1, \hdots, n\}$, due to \cref{eq:common_joint_dist} and the independence of the observation noise. Note the expectation is taken over the observation noise and latent randomness. Finally, averaging over the corpus $\frac{1}{n-1} \sum_{i=1}^n (\mathbf{\hat{\Delta}}_{i,1}-\frac{1}{n} \sum_{i=1}^n \mathbf{\hat{\Delta}}_{i,1}) (\mathbf{\hat{\Delta}}_{i,2}^\top - \frac{1}{n} \sum_{i=1}^n \mathbf{\hat{\Delta}}_{i,2}^\top)$ then provides an unbiased estimate of $\bm{\Lambda}$. The reduction in efficiency due to sample-splitting for this estimate within each RCT can also be partially mitigated through cross-fitting techniques \citep{doubleml} or a jackknife approach as explored in \citet[Section 4.2]{bibaut2024learning}.  In our work we find the hierarchical modeling approach to be natural as it incorporates the observed heteroscedasticity in observation noise across RCTs which vary significantly in size in our corpus.

\section{Proxy Score and Sensitivity}
\label{app:proxy_score}

\begin{figure}[!htb]
    \centering
    \includegraphics[width = 0.9\linewidth]{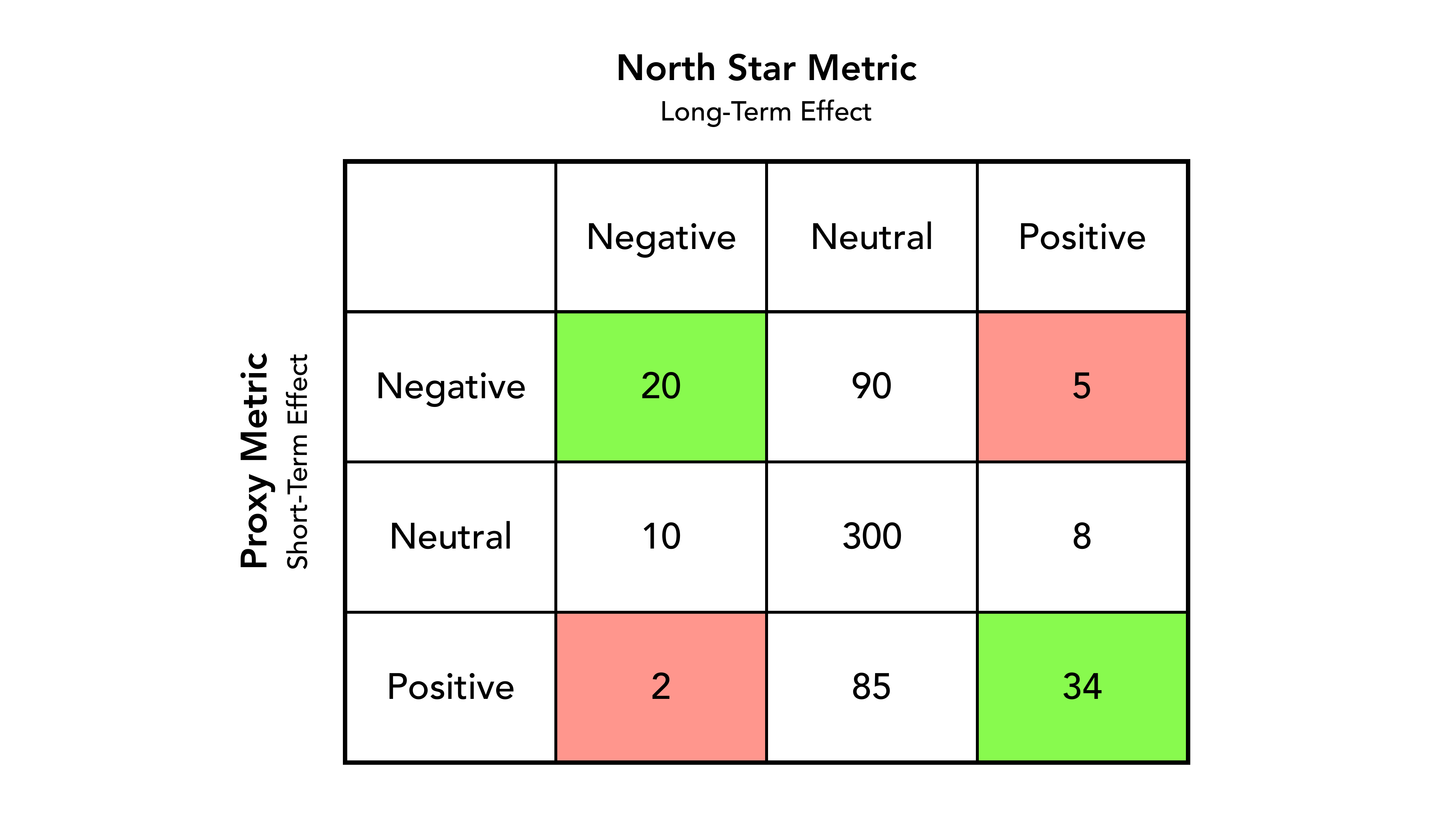}
    \caption{A synthetic contingency table which captures the alignment of the decisions induced by the t-statistics of the TEs of the north star metric and a proxy metric. }
    \label{fig:proxyscore}
\end{figure}

Here we explain several performance criterion we use for proxy metrics which are further detailed in the literature \citep{richardson2023pareto}. In order to define both quantities, recall that a TE metric is often used to make a downstream decision by thresholding its t-statistic, $\tstat = \frac{\hat{\Delta}}{\sqrt{\Var(\hat{\Delta})}}$ as $\tstat > 2 \to + \text{ (positive)}$, $-2 < \tstat < 2 \to 0 \text{ (neutral)}$, and $\tstat < -2 \to - \text{ (negative)}$. The formal definitions of proxy score and sensitivity are most easily defined in the context of a  contingency table visualized in \cref{fig:proxyscore} which takes these decisions as inputs. The contingency table tabulates the  decisions induced by a particular observed short term proxy metric and the observed long-term north star metric jointly over 554 A/B tests.

The green cells in \cref{fig:proxyscore} represent cases where the proxy and long-term north star are both statistically significant and move in the same direction (i.e. \emph{Detections}). The red cells in \cref{fig:proxyscore} again represent cases where the proxy and long-term north star are both statistically significant, but where proxy and long-term north star are misaligned (i.e. \emph{Mistakes}). The remaining cells correspond to cases where at least one of the metrics is not statistically significant. The relative importance of these cells is more ambiguous.

In this setting, the sensitivity can be defined as:
\begin{align}
    \text{Metric Sensitivity} = \frac{\text{Num. of expts. proxy is significant}}{\text{Num. of total expts.}}. \nonumber
\end{align}
Here the numerator can be obtained by summing over the first and last rows of the table.
The proxy score can be defined as,

\begin{align}
    \text{Proxy Score} = \frac{\text{Detections} - \text{Mistakes}}{\text{Num. expts. long-term north star is significant}}. \nonumber
\end{align}

The denominator here can be obtained by summing over the first column and last column. 
The sensitivity metric captures the ability of a metric to detect a statistically significant effect -- which inherently takes into account its inherent moveability and susceptibility to experimental noise. Given that north star metrics are often noisy and slow to react in the short-term the goal of a proxy is to be sensitive.

The proxy score rewards metrics that are both sensitive and directionally aligned with the north star. Sensitive metrics need only populate the first and third rows of the contingency table. However, metrics in the first and third rows can only increase the proxy score if they are in the same direction as the long-term north star. A similar score, called \emph{Label Agreement}, has been used by \cite{dmitriev2016measuring}. 
    It is worth noting, these measures are still imperfect in the sense that comparisons are made against the decisions induced by the noisy estimate of the long-term outcome TE not the latent population long-term outcome TE. This is further complicated by the fact that we empirically find that experimental noise in the A/B tests is correlated between short-term proxies and the long-term outcome (i.e. the phenomena detailed in \cref{fig:denoising}). In fact, this phenomena provides partial motivation for our definition of denoised proxy quality.

\bibliographystyle{plainnat}
\bibliography{references}
\end{document}